\documentstyle[12pt]{article}
\begin{document}
\begin{center}

{\ Natural Color Transparency in High
Energy (p,pp) Reactions}\vspace{\fill}\\
B.K.~Jennings\\
{\em TRIUMF, 4004 Wesbrook Mall, Vancouver, B.C., V6T 2A3, Canada}\\
and\\
G.A.~Miller\\
{\em Department of Physics, FM-15, University of Washington, Seattle, WA 98195,
   USA}
\vspace{\fill}\\
{\bf ABSTRACT}
\end{center}
New parameter free calculations including a variety of necessary
kinematic and dynamic effects
show that the
results of BNL
$(p,2p)$ measurements are consistent with the expectations
of color transparency.
\vspace{\fill}
\begin{center}
  (submitted to Phys. Lett. B)
\end{center}
\vspace*{\fill}

\pagebreak

The anomalously large transmission of a hadron in the nuclear medium
following or preceding a hard interaction is commonly referred to as
color transparency\cite{mueller,brodsky}.
  This topic has been actively studied theoretically
\cite{farrar}-\cite{jennings2}. 
Color transparency depends on the formation of a small-sized wavepacket
or point-like configuration PLC in
a high momentum transfer quasi-elastic reaction.
The effects of color screening or color
neutrality \cite{low} suppress the interaction between the PLC and
the nuclear medium. At the present experimental
energies the  PLC expands as it moves through the nucleus and therefore
does interact. This expansion is treated in
Refs.\cite{farrar,jennings,jennings2,jmplb,jmprl}.

Color transparency is under
active experimental investigation
at BNL \cite{carroll,sh2} and at SLAC\cite{milner,bobm}.
The $(p,2p)$ experiment of
Carroll et al\cite{carroll} finds  a transparency $T$ (ratio of nuclear to
hydrogen cross section per nucleon after removing the effects of
nucleon motion) with an oscillatory pattern. $T$ increases as the
beam momentum increases from 5 to 10 $GeV/c$ but
then decreases.  The SLAC experiment is unpublished at this time, but
a preliminary report \cite {bobm} found no effect of color transparency (CT)
in (e,e'p) reactions for
$Q^2$ between 1 and 7 GeV$^2$. However, the stated systematic errors are about
$\pm$ 10\% which are  large compared to the size of the predicted CT effects.

The purpose of this note is to include all of the
known necessary kinematic and dynamic effects:
a  proper treatment of the longitudinal
(parallel to the virtual photon
or incident proton momentum)  component of the momentum of the
 detected protons\cite{kj,fsz};
 computing the measured experimental observable which involves an
integration over the transverse momentum of the struck proton;
 including the effects on interference between PLC and non-PLC
configurations produced in proton-proton elastic scattering
\cite{brodsky2,ralston}; and, a realistic treatment of
the baryonic components of the PLC\cite{jmprl}.
Thus our aim here is to show that including the three kinematic and
dynamic effects mentioned above along with the color transparency of
ref.~\cite{jmprl} leads to a natural explanation of the existing BNL data.
No free parameters are used in the present work.

We start the analysis by recalling the mechanisms proposed
to understand the (p,pp) reaction. The  suggestion of
Ralston and Pire \cite{ralston}, is that the energy
dependence is caused by an interference between a hard amplitude which produces
a PLC, and a soft one (blob-like
configuration BLC) which does not. The
Ralston and Pire idea is that
 the BLC is due to the Landshoff process\cite{pire}.
Another mechanism with a similar effect
is that of Brodsky and de Teramond \cite{brodsky2}
 in which the two-baryon system couples to charmed quarks
(there is a small (6q) and a BLC which is a (6q,$c\bar c$) object)
The Brodsky-de Teramond idea is motivated
by the fact that the mass scale of the rapid energy
variation in
 $A_{NN}$\cite{kirs}  and in the measured transparency  matches that
of the
charm threshold.

Both of the mechanisms of \cite{ralston} and\cite{brodsky2}
are well motivated, but imprecisely understood. For example,
the nature of the Landshoff term is uncertain. At high energies one
expects that this component is suppressed by a Sudakov factor \cite{mueller2}.
At present energies, the energy dependence and phase of the term are not well
determined, and the size of the object produced by the
Landshoff process is not well known.
See e.g. ref. \cite{botts}.
Nevertheless, the Sudakov effects can be expected to
a set of configurations with a range of different  sizes.
Similarly even if
the details of the $c\bar c$ production amplitudes are not yet well-
established,
threshold effects will naturally lead to mixtures of BLC and PLC.
Thus, it's natural to discuss high Q$^2$ elastic proton-proton scattering
in terms of configurations of different sizes.
Separating the contributing configurations into two, a PLC and a BLC
is only a simple first step.

The two ideas about the BLC can be used to qualitatively explain the
oscillations observed in the (p,pp) data, but do not quantitatively reproduce
features of the data Ref. \cite {kopel,jmplb} when combined with a proper
treatment of the expansion\cite{jmplb}.  The agreement is not quantitative
because including the non-zero absorption of the expanding PLC and the
non-complete absorption of the BLC tends to make the two terms similar and
weakens the interference effects.

Another relevant effect is to properly account for the correct momentum of
the detected protons. Consider for example, the (e,e'p) reaction
in which the virtual photon has the momentum $\vec q$. Suppose the
detected proton has momentum $\vec p$ with $\vec p=\vec q +\vec k$.
For quasi-elastic kinematics $\vec p=\vec q$ and $\vec k =\vec 0$.
Fermi motion or final state interactions can lead to a non-zero value
of $\vec k$. This is old news. However the  non-zero
components of $\vec k$ in the direction of $\vec q$ ($k_z$)
have been shown to give a large numerical effect \cite{kj}. A detailed study
 of this
``Fermi motion" effect was made by Frankfurt et al \cite{fsz}.
The momentum $\vec k$ is sometimes called the momentum of the struck
nucleon (and is that quantity if the plane wave Born approximation is valid
). We also use the term ``momentum of the struck proton", but only as an
abbreviation.
For the (p,pp) reaction the momentum of the detected protons can also be
described in terms of the momentum of the struck nucleon \cite{carroll}.
Indeed, the data are presented in terms of bins of $k_z$, the component of
$\vec k$  parallel to the beam direction.
This is especially important for the (p,pp) reaction because the proton-
proton scattering cross section varies approximately as $s^{-10}$ and here
$s=2 M^2 +2 M E_{LAB} -2 k_z P_{LAB}\approx 2 M P_{LAB}(1-k_z/M)$
where M is the
proton mass.
Ref \cite{fsz}  examined the (p,pp) reaction,
including the effects of non-zero values of $k_z$ in their calculation, and
obtained qualitative agreement with the 10 GeV data, but not with data
taken at the other energies. Agreement was also obtained by
Kopeliovich\cite{boris} for some energies.
Neither group included any effects of
interference between PLC and BLC.

Our calculation of the nuclear (p,pp) cross sections uses the
quantum mechanical treatment of  color transparency developed in
ref.~\cite{jennings,jennings2,jmprl}.
In that approach the time for the PLC to expand depends on
the masses of the baryonic components of the PLC. In early work
\cite{jennings,jennings2} only
a single average state of mass $M_1$ was used. Then the
time for the PLC to expand is ${2P_{LAB}\over (M+M_1)} {1\over (M_1-M)}$.
Recently we \cite{jmprl} included the effects of the continuous baryon
spectrum by
using  measured proton diffractive dissociation
and electron deep inelastic scattering data to constrain the baryon masses.
Our discussion of the formalism shall be brief and concentrate on the newer
kinematic and  dynamic aspects. For details see Refs.
\cite{jennings2,jmplb,jmprl}.

To be definite, consider the (e,e'p)
 high momentum transfer process in which a photon of
three-momentum $\vec{q}$ is absorbed and a nucleon of momentum $\vec{p}$ leaves
the nucleus. As usual, $q^2=-Q^2$.
Consider knockout from only a single shell model orbital, denoted by $\alpha$.
The observable cross sections are computed by making
incoherent sums over all the occupied orbitals.

Let the amplitude be defined as $\cal M_{\alpha}$, which can be described
in terms of a series involving multiple wavepacket (PLC) -nucleon
scatterings.
The first term is the plane wave  Born term $B_{\alpha}$ in which the
wavepacket undergoes no interactions.
If full color transparency is obtained, this is the only
term to survive. The first correction to the Born term is the scattering
term or second term denoted by
$ST_{\alpha}$. To the stated order we have:
${\cal M}_\alpha=B_\alpha+ST_\alpha
$
. The Born term is given by
\begin{eqnarray}
B_\alpha=\langle\vec p|T_H(Q)|\alpha\rangle = F(Q^2)\langle\vec p-\vec q|
\alpha\rangle,
\end{eqnarray}
in which $T_H(Q^2)$ is the hard scattering operator.
Specific effects of spin are  ignored here and
throughout this work.

The second term is denoted by
\begin{eqnarray}
ST_\alpha = \langle\vec p|U\;G\;T_H(Q)|\alpha\rangle,
\end{eqnarray}
where $G$ is the Green's operator for the emerging small object (PLC).
This object
is a wave packet of which the nucleon is just one
component.
The operator $U$ represents the interaction between the ejected baryon and the
nuclear medium. The nuclear interaction can change the momentum of the ejectile
and also excite or de-excite the internal degrees of freedom.
Our approach is to treat the Green's function G as a sum of baryonic
propagators each
denoted by $m$.
The eikonal Green's function for the emerging baryon  is:
\begin{eqnarray}
G(Z,Z') = \sum_m G_m(Z,Z')= \sum_m{\theta(Z-Z')e^{ip_m(Z-Z')}\over 2ip_m},
\end{eqnarray}
with
$\vec p^{\;2}_m=\vec p^{\;2}+M^2-M^2_m.$
Here $M$ is the nucleon mass, $M_m$ the mass of the baryonic state. Our
notation is that
$p = |\vec{p}| $ and $p_m=|\vec {p_m}|$.
Terms beyond the first order in  $U$ are included
by exponentiation, which is an excellent approximation for our
applications \cite{greenberg}.

The work of ref. \cite{jennings2} involves a simple model in which
the operator $U$ gives only one excited state (m=1) when acting on a nucleon.
In that model
\begin{eqnarray}
ST_\alpha =-F(Q^2)\int d^2BdZ\rho(B,Z)\;{e^{-ipZ}\over (2\pi)^{3/2}}
\int
^Z_{-\infty} dZ'e^{-ip(Z'-Z)}{\sigma_{eff}\over 2} (Z,Z')e^{iqZ'}\langle
\vec B,Z'|\alpha\rangle,    \label{eq:st1}
\end{eqnarray}
where the $\rho( B,Z)$ is the nuclear density, and
\begin{eqnarray}
\sigma_{eff}(Z,Z')\equiv\sigma\left(1-e^{i(p-p_1)(Z'-Z)}\right)
,\label{eq:st2}
\end{eqnarray}
and $\sigma$ is the proton-nucleon total cross section.
This is eq (32) of ref. \cite{jennings2}.

Now we can exhibit the importance of the momentum $(k_z)$ of the struck
nucleon.
We take $p=q+k_z$ for components parallel to the direction of $\vec q$.
Change the integration variable $Z^\prime$ to $D$ via
$Z^\prime=Z-D$. Then
\begin{eqnarray}
ST_\alpha =-F(Q^2)\int d^2BdZ\sigma \rho(B,Z)\;{e^{-ik_z Z}\over (2\pi)^{3/2}}
\int
^{\infty}_0 d D e^{i k_z D }\left(1-e^{-i(p-p_1)D}\right)\langle
\vec B,Z'|\alpha\rangle.
\end{eqnarray}
The effect
of $k_z$ appears in two places. The $e^{-ik_z Z}$ factor is the same as
in standard Born or Distorted Wave Born calculations (DWBA or Glauber
optical model).
The $e^{i k_z D }$ factor involves a modification of the color
transparency  physics.
To see this note that
the real part of $ST_\alpha$ dominates the numerics.  Then
$cos(k_zD)-cos\left((k_z-p+p_1)D\right)\approx {D^2\over 2}
(p-p_1)\left(p-p_1 -2 k_z\right)$ for small $D$.
Since $p>p_1$
a  positive value of $k_z$ reduces this term and increases the transparency for
any given value of p.
This latter effect does not occur in the color transparency of ref.
\cite{farrar}, but would occur in models in which the baryon-nucleon
interaction is treated as a finite-dimensional matrix\cite{boris,sumrule}.

The above paragraph
is meant as a simple explanation of a numerical effect that is
surprisingly large.
The computations of $(p,pp)$ reactions are more involved since there is one
incident proton wave function and two outgoing ones. Furthermore the
amplitudes for the production of both the PLC and BLC must be taken into
account.  However the qualitative
effect of including non-zero  values of $k_z$ (here z is the direction of
the beam proton) is similar to the electron
scattering case. Note also that  our more realistic calculations \cite{jmprl}
replace the single mass $M_1$ by an appropriate distribution of masses.

The next step is to discuss the observables measured in ref.\cite{carroll}.
Let the four-momentum of the target proton be denoted as $(M,\vec k)$.
The transparency T is defined as a ratio $T=d\sigma/d\sigma^B$ with

\begin{equation}
d\sigma =\int_{k_a}^{k_b}dk_z \int\int dk_x dk_y (d\sigma^A/dt)(s)
\label{eq:bin1}
\end{equation}
and
\begin{equation}
d\sigma^B=\int_{k_a}^{k_b}dk_z \int\int dk_x dk_y (d\sigma^B/dt)(s)
\label{eq:bin2}
\end {equation}
in which the superscript A denotes the nuclear cross section divided by the
number of target protons and the subscript B denotes the same quantity but
computed in the plane wave Born approximation. In the experiment the
integrations over the transverse momenta $k_x,k_y$ were limited to about
250 MeV/c. This corresponds to almost all of the probability so we
integrate over all $k_x,k_y$ in our calculations. The integration over the
transverse components of $\vec k$ reduces the Glauber DWBA result for T
by about 30\%.
The data of ref.~\cite{carroll}
are presented in terms of T for each $(k_a,k_b)$.

We now present the results. The calculations for an Al target
(three beam momenta and four bins of $k_z$)
are shown in Fig 1.
The solid curves show the effect of CT using the complete power law
form of the distribution
of baryon masses as in ref. \cite{jmprl} and dashed curves
show the results of using Ref.\cite{jennings2}
with a value of $M_1=1550 MeV$. This value gives small enough
color transparency at low $Q^2$ so as to be consistent with the NE-18 data.
The Ralston-Pire parameterization of $d\sigma/dt $ for the free protons is
used here along with their separation of the  BLC and PLC terms.
Details of the implementation of this are to be found in ref \cite{jmplb}.
The use of the corresponding Brodsky-de Teramond model for the PLC-BLC
interference would lead to similar results for the energy range we consider
here \cite{jmplb,jmprl}.
The data are from ref.~\cite{carroll}. The target-dependent and target
independent
uncertainties in the normalization of T are about 10\% and 25\%
respectively \cite{carroll}.
 We multiply the central values of the Al, Cu, and Pb
data points  by a factor of 0.75, and those
of Li and C by 0.85. This is consistent with the published errors.
Each data ``point" represents a bin of $k_z$,  represented as a
horizontal line.
The integration of Eqs. (\ref{eq:bin1}),(\ref{eq:bin2})
over the small bins 0.1 or 0.2 GeV/c of ref.\cite{carroll}
causes negligible differences with simply  using
the central values.
The use of a distribution of masses starting at $M+m_{\pi}$
increases the computed $d\sigma/d\sigma_B$ at the lowest beam momentum.
In either case, including the effects of color transparency
gives good agreement with the data.

Fig. 2, shows a summary of the different calculations for $^{27}$Al. Here as in
Figs.  1 and 3 the solid curve represents the full color transparency
calculation, the dashed curves are obtained by neglecting the effects of the
PLC-BLC interference, and the dotted curves represent the use of the standard
optical model or ``Glauber" calculation. These latter curves fall far below the
data, but including the effects of CT leads to a reasonable reproduction of the
data.

Fig. 3 shows the A-dependence of the data of ref.\cite{carroll}.
These are taken for a bin of $k_z$ ranging from
-0.2 GeV/c to 0.1 GeV/c.  Here the use of Eqs. (\ref{eq:bin1}) and
(\ref{eq:bin2})
does matter. To see this compare the $^{27}$Al data of this figure with the
previous figures.
The Glauber standard optical model leads to
results (dotted curves) that again fall well below the data, this time
 for each target nucleus.
The solid curves, which show our full calculation, are in excellent
agreement with all of the data, except for the 12 GeV Al data point.
The dashed curves
show the results obtained without the BLC-PLC interference
Ralston-Pire effect, so we see that
the latter helps to account for the energy dependence, even though it
is not a very large numerical effect.

It is necessary to comment on the
single particle nuclear shell
model wave functions used here.
Harmonic oscillator ($\hbar\omega=41MeV/A^{1/3}$ ) wave functions are used
for light nuclei (Li and C). The other nuclei are treated in
the Hartree-Fock (HF) approximation with
the SGII interaction of Ref.~\cite {hf}. (HF wave
functions are used in obtaining the results of Figs.~1,2.)
If the
oscillator  frequency is chosen appropriately,
the HF wave functions are well approximated by a single harmonic
oscillator wave function.
 Thus we find
the most important effect of using the HF wavefunctions is to shift
the value of $\hbar\omega$ from $41 MeV/A^{1/3}$  to one that more precisely
represents the nuclear mean square radius. The effect of using
HF wave functions  with the correct exponential dependence at long
distance on $d\sigma/d\sigma^B$ is
largest, a 20\% increase independent of energy, for the Glauber optical
model calculations for the Pb target. But the effect is negligible
for the CT  calculations. See also ref.\cite{lee}.

The main effect of the color transparency is to provide an increase in the
predicted magnitude of the cross sections. One might wonder if some
combination of reasonable effects applied to the Glauber optical model
calculations yield enhancements that lead to reproducing the data without
including color transparency effects. One possibility is to to claim
\cite{Yazaki} that the experimental resolution allows the inclusion of all
nuclear excited states in the measurement. In that case, the optical
model wave functions would be computed using the proton-proton reaction cross
section $\sigma_r$ instead
of the total cross section $\sigma$ as we have done. However,
$\sigma_r/\sigma\approx 0.85$  \cite{pdg}
and has
only a small energy dependence, for  the energy range of interest here.
Calculations show that including
this effect provides only 10-20 \% enhancements. Including the effects of
nucleon-nucleon short range correlations leads to a 20\% increase for
$^{12}C$ \cite{lee}, but is a much smaller effect for heavier nuclei.
Thus, in the absence of CT effects,  we see no possibility  to elevate the
the DWBA cross sections
to the levels observed by the experiment.

We believe the BNL experiment and the present calculations calibrate the size
of color transparency effects. Thus we determine the energies for
measurable effects. Our  calculations provide a good
guide  to the (e,e'p) experiments. The SLAC Ne-18 experiment is set
for very small values of $k_z$, so
 the previous results of ref \cite{jmprl} stand
unchanged (except possibly for the effects of using improved nuclear structure
information for heavy targets). Since the predicted transparency is not large
in the range of $Q^2$ between 1 and 7 GeV$^2$, we expect that our earlier
calculations of color transparency effects for the power-law form of $g(M_X^2)$
will  not be ruled out by the final results of the
SLAC experiment\cite{milner,bobm}. Furthermore, our CT
calculations predict enhancements for $Q^2$ between 7 and 15
GeV$^2$. Thus a higher energy experiment which examines the $k_z$ dependence
should observe measurable effects.

Thus, we summarize:
If CT effects are included,
the  qualitative features of
the data of ref \cite{carroll} can be reproduced in a very natural way.
No adjustment of parameters  is needed. If the new BNL (p,pp)
color transparency
experiment \cite{sh2} confirms the central values of the
older results, one can be confident that color
transparency has been discovered.

Acknowledgment: One of the authors (BKJ) thanks the Natural Sciences and
Engineering Research Council of Canada for financial support.
The other (GAM) thanks the DOE for partial support. Discussions with
W.R.~Greenberg and M.~Strikman
are appreciated.  We thank G.~Bertsch
for making the HF program of ref.~\cite{hf} accessible to us.

\pagebreak

\begin{center}
Figure Captions
\end{center}

Figure 1: Effect of using  distributed baryonic masses.
Solid uses  distributed masses. Dashed- $M_1$ =1550 MeV.
 The data in all figures are those of Ref.~\cite{carroll}.

Figure 2. Full Al data.
Solid- full calculation.
Dashed- with out the Ralston-Pire effect. Dotted -Glauber.

Figure 3.  A dependence of transparency.
Solid- full calculation.
Dashed- with out the Ralston-Pire effect. Dotted -Glauber.

\end{document}